\documentclass[pra,twocolumn,showpacs,superscriptaddress]{revtex4}
\usepackage{amsfonts}
\usepackage{amsmath}
\usepackage{amssymb}
\usepackage{graphicx}

\newcommand{\be}{\begin{eqnarray}}
\newcommand{\ee}{\end{eqnarray}}

\begin{document}

\title{Smooth cut-off formulation of hierarchical reference theory 
for a scalar $\phi^4$ field theory}
\author{Cristian D. Ionescu\footnote{Member on leave of
absence of the INFPLR-ISS, P. O. Box MG-23, RO 76911 Bucharest, Romania}}
\affiliation{Dipartimento di Fisica, Universit\`a degli Studi di Milano, 
Via Celoria 16, 20133 Milano, Italy}
\author{Alberto Parola}
\affiliation{Dipartimento di Fisica e Matematica, Universit\`a dell'Insubria, 
Via Valleggio 11, 22100 Como, Italy}
\author{Davide Pini}
\affiliation{Dipartimento di Fisica, Universit\`a degli Studi di Milano, 
Via Celoria 16, 20133 Milano, Italy}
\author{Luciano Reatto}
\affiliation{Dipartimento di Fisica, Universit\`a degli Studi di Milano, 
Via Celoria 16, 20133 Milano, Italy}



\begin{abstract}
The $\phi^4$ scalar field theory in three dimensions, prototype for the study of 
phase transitions, is investigated 
by means of the Hierarchical Reference Theory (HRT) in its smooth cut-off formulation. 
The critical behavior is described by scaling laws and critical exponents
which compare favorably with the known values of the Ising universality class. The 
inverse susceptibility vanishes identically inside the coexistence curve, providing
a first principle implementation of the Maxwell construction, and shows the expected 
discontinuity across the phase boundary, at variance 
with the usual sharp cut-off implementation of HRT. The correct description of
first {\it and} second order phase transitions within a microscopic, non-perturbative 
approach is thus achieved in the smooth cut-off HRT.  
\end{abstract}

\maketitle

\section{Introduction}
The equilibrium properties of classical fluids have been extensively studied 
in the past decades and very successful theories of both the gas and 
the liquid states are available by now {\cite{Donald,caccamo}}.
However, a fully satisfactory description of the liquid-vapor 
phase transition within a microscopic liquid state approach has still to come.
Although the qualitative 
features and the universal properties characterizing the transition 
are known since the development of the Renormalization Group (RG) approach \cite{rg}, 
a quantitative theory able to predict phase boundaries for specific fluid models 
is not avaliable yet: 
Most of the theories developed in liquid-state physics deal mainly
with short-range correlations and short-wavelength
density fluctuations. That is why they are not able to reproduce, even
qualitatively, the phenomenology of the critical region.
Furthermore, these liquid-state theories give an unsatisfactory
description of the thermodynamics along the first order 
liquid-vapor transition line. The convexity of the free energy is not
guaranteed in the integral equation approach or in mean-field theories
and in fact is always violated inside the two-phase region. This intrinsic
deficiency is usually overcome by means of {\sl ad hoc} procedures, 
like Maxwell double tangent construction.
An accurate description of the long-wavelength fluctuations
is implemented within the liquid-state framework in the Hierarchical Reference Theory of
fluids (HRT) \cite{pra,mare}, where the basic RG concept of selective turning on
of fluctuations is built into a genuine liquid-state theory. This feature
made HRT the only practical scheme able to determine
both universal and non-universal critical properties of fluids, so far. 
Moreover, the proper treatment of long-wavelength fluctuations forces 
the free energy to display the correct convexity also in the two-phase region,
thereby incorporating Maxwell construction: A major achievement of HRT.
Although quite successful, the current sharp cut-off formulation of HRT 
predicts a diverging compressibility along the binodal (in any dimension smaller than four)
\cite{max} while, for a scalar order parameter, a finite value of the 
inverse compressibility at coexistence is expected. 
In this paper we show that this failure of the HRT approach in the sharp cut-off formulation
can be overcome by studying the {\sl smooth cut-off} implementation, already introduced several
years ago \cite{smooth}. In order to clarify the qualitative properties expected within 
this approach, we consider the simplest model displaying a phase transition within the
Ising universality class: The scalar $\phi^4$ field theory in three dimensions.
A somehow related formalism has been recently put forward by Caillol {\cite{Cai}} 
who developed a non-perturbative renormalization-group theory for fluids, within the Grand 
Canonical ensemble and applied it to the Kac model. 
 
The paper is organized as follows: 
In Section II we review the sharp and the smooth cut-off formalism 
for a general microscopic fluid model. In Section III this 
formalism is applied to a $\phi^4$ field theory, where several simplifications occur. 
An approximate closure relation is introduced and discussed. 
The numerical solution of the equations at low temperatures and the
comparison between the two approaches is presented in Section IV, where 
the behavior at coexistence is analyzed. The critical properties of HRT in the
smooth cut-off formulation are derived in Section V, while Section VI contains a short 
summary of the results and some perspective.

\section{HRT equations in sharp and smooth cut-off formulations} 
The common principle underlying HRT, in its various formulations, is 
the selective treatment of density fluctuations on different length scales.
Starting from a mean field approximation, the effects of density fluctuations 
on short wavelengths are included first. This is accomplished by
turning on the attractive part of the potential progressively,
starting from its Fourier components with larger wave vectors.
A key role is therefore played by the cut-off wave vector $Q$ 
which separates the Fourier components already taken into account (those
with $k > Q$) from those ignored ($k < Q$). The full spectrum of fluctuations
is then included only in the $Q \to 0$ limit, while at $Q\to\infty$ the mean field
approximation is recovered. The variation in the free energy induced 
by an infinitesimal change in $Q$ can be computed exactly by use of perturbation
theory, leading to a differential ``evolution" equation for the thermodynamics
of the system. 

We study a system of particles in dimension $d$, interacting via a two-body potential $v(r)$ 
formally written as the sum of two contributions:
\begin{equation}
v(r)=v_R(r)+w(r) 
\label{2bodyp}
\end{equation}
where $v_R(r)$ is a short-range, repulsive reference part, and $w(r)$
is a mostly attractive term 
defined by a regular function which can be Fourier transformed. 
The properties of the system interacting via $v_R$ are assumed to be known by
other liquid state theories or simulations. 
In order to implement the principle of HRT, the Fourier components of $w(r)$ have
to be included gradually. Therefore we define a sequence of $Q$ systems characterized
by an interaction $v_Q(r)$:
\begin{equation}
v_Q(r)=v_R(r)+w_Q(r)
\label{2bodypQ}
\end{equation}
interpolating between the reference system (obtained for $w_Q(r)=0$) and the fully interacting 
model (for $w_Q(r)=w(r)$). The difference between sharp and smooth cut-off formulation
resides in the way the cut off on the Fourier components of the potential
is implemented, as discussed in the following.

\subsection{Sharp cut-off}

In the sharp cut-off formulation, the precise definition 
of $w_Q$ is conveniently given in Fourier space: 
\begin{eqnarray}
\tilde w_Q(k)=\begin{cases}
\tilde w (k) & \mbox{for } k>Q\\
0 & \mbox{for } k<Q
\end {cases}
\label{sharptail}
\end{eqnarray}
where the tilde denotes Fourier transform.
Here, the cut-off $Q$ varies between $Q=\infty$, where the potential reduces
to the reference part, and $Q=0$, where the physical interaction (\ref{2bodyp}) is recovered. 
Due to the long range oscillating tail in $w_Q(r)$, induced by the sharp cut-off, no
phase transition occurs in each $Q$-system until the $Q\to 0$ limit is attained. 
It is now possible to write a differential equation for the evolution of the 
free energy $A_Q$ of the $Q$-system when the cut-off is varied. This equation,
though exact, is not closed because it involves the two body correlation
function of the $Q$-system, and suitable approximations linking thermodynamics 
and correlations must be introduced. Before discussing the adopted closure, 
following the RG approach, it is convenient to introduce an additional definition
in order to take into account the Fourier component of the
potential with vanishing wave vector since the very beginning of the 
integration procedure. The $k=0$ term in fact 
gives rise to the mean-field contribution to the free energy which 
provides a physically correct starting point for the description of phase 
transitions. This is achieved by introducing a {\sl modified} free energy density defined as: 
\begin{eqnarray}
{\cal{A}}_Q=-\frac{\beta}{V}A_Q-\frac{1}{2}\rho\left[\phi(r=0)-\phi_Q(r=0)\right]+\nonumber\\
\frac{\rho^2}{2}\left[\tilde{\phi}(k=0)-\tilde{\phi}_Q(k=0)\right] 
\label{sharpen}
\end{eqnarray}
where $\phi(r)=-\beta w(r)$, $\beta=1/k_BT$ and $\rho$ is the particle density.
Analogously, the modified direct correlation function includes the Random Phase Approximation
(RPA) contribution:
\begin{equation}
{\cal{C}}_{Q}(k)\equiv \tilde{c}_Q(k)+\tilde{\phi}(k)-\tilde{\phi}_Q(k)
\label{sharpcorel}
\end{equation}
where $c_Q(r)$ is the usual direct correlation function \cite{Donald} of the $Q$-system
(with the inclusion of the ideal gas term).
The evolution of the (modified) free energy satisfies the exact differential equation:
\begin{equation}
\frac{d{\cal{A}}}{dQ}=-\frac{d}{2}\,\Omega_d \,Q^{d-1}\ln
\left(1-\frac{\tilde{\phi}(Q)}{{\cal{C}}_Q(Q)}\right)
\label{sharpevol}
\end{equation}
where $\Omega_d$ is the volume of the unit sphere in $d$ dimensions divided by $(2\pi)^d$.  
An analogous equation governs the evolution of ${\cal{C}}_Q(k)$ which, however,
depends on the three and four body correlation functions of the $Q$-system, thereby 
giving rise to an exact hierarchy of differential equations. In the following 
Section we will provide an approximate closure at the level of 
the first equation of the hierarchy, Eq. (\ref{sharpevol}). 

\subsection{Smooth cut-off}

In the smooth cut-off formulation of HRT \cite{smooth} 
the discontinuity in the Fourier transform of $w_Q$ induced by the definition 
(\ref{sharptail}) is removed by the alternative choice
\begin{equation}
\tilde w_t(k)=\tilde w(k)-e^{-2t}\tilde w(k\,e^{t})
\label{smoothtail}
\end{equation}
which identically vanishes at $t=0$ and tends to $\tilde w(k)$ for $t\to\infty$. 
If $\tilde w(k)$ varies on the characteristic wave vector scale $\Lambda$, the quantity
$Q=\Lambda e^{-t}$ acts as an effective cut-off wave vector: The Fourier components 
of $\tilde w_t(k)$ for $k < Q$ are efficiently suppressed and the long range 
repulsive tail present in $w_t(r)$ contrasts the tendency towards phase separation
at any finite $t$. 
The sequence of intermediate potentials $\tilde w_t(k)$ defined in (\ref{smoothtail})
belongs to the class studied in Ref. \cite{Cai} and corresponds to the particular choice of cut-off function
$R_t(k)\,\tilde w(k)=e^{-2t} \tilde w(k\,e^t)\, [\tilde w(k)-e^{-2t}\tilde w(k\,e^{t})]^{-1}$.
The requirements stated in \cite{Cai} are satisfied by $R_t(k)$ provided $\tilde w(k)$
decays rapidly at large wave vectors.
First order perturbation theory provides the
change in the free energy induced by an infinitesimal increase of the parameter $t$.
By use of the same the definition (\ref{sharpen}) already introduced,
the exact evolution equation of the (modified) free energy in the smooth cut-off formulation 
of HRT reads:
\begin{equation}
\frac{d{\cal A}_t}{dt} =\frac{1}{2}\int \frac{d^dk}{(2\pi)^d}
F_t(k) \frac{d\tilde\phi_t(k)}{dt}
\label{hrt}
\end{equation}
where $\tilde\phi_t(k)=-\beta\tilde w_t(k)$ and the two point function 
$F_t(k)=-\left [\tilde c_t(k)\right ]^{-1}$ 
is just $\rho$ times the structure factor $S_Q(k)$ of the $Q$-system. 
As usual, it is more convenient to introduce 
the modified direct correlation function ${\cal C}_t$ (Eq. (\ref{sharpcorel}))
which includes the total interparticle potential in mean-field approximation,
because we expect that its convergence properties as $t\to\infty$ will be better than 
those of the ``bare" correlations $c_t(k)$. By use of Eqs. (\ref{sharpcorel}) and 
(\ref{smoothtail}) 
we get a formal expression for the bare two point function appearing in Eq. (\ref{hrt}):
\begin{eqnarray}
F_t(k)=-\left [ {\cal C}_t(k) -\tilde\phi(k)+\tilde\phi_t(k)\right ] ^{-1}\nonumber\\
=-\left [ {\cal C}_t(k)-e^{-2t}\tilde\phi(k\,e^t)\right ]^{-1} 
\label{fk}
\end{eqnarray}
Analogously to the sharp cut-off case, the flow equation (\ref{hrt}) 
for ${\cal{A}}_t$  is not closed and a suitable approximation for 
${\cal C}_t(k)$ must be introduced.

\section{HRT equations for the $\phi^4$ scalar field theory}

In order to test the accuracy of HRT in the description of phase transitions, 
we specialize the general equations (\ref{sharpevol}) and (\ref{hrt})
to a particular model of fluid. We first note that, in systems with regular short-range 
attractive interactions, the Fourier transform of $\phi(r)$ is a positive, monotonic decreasing 
function of the wave vector, quadratic in $k$ at small $k$. The simplest model 
with these features is defined by an effective interaction of the form:
\begin{equation}
\tilde\phi(k)=\begin{cases}
b\,(\Lambda^2-k^2) & \mbox{for } k < \Lambda \\
0 & \mbox{for } k > \Lambda
\quad
\end{cases}
\label{phi}
\end{equation}
By substituting this form into Eqs. (\ref{sharpevol}) and (\ref{hrt})
we get, respectively
\begin{eqnarray}
\label{parasharp}
\frac{d{\cal{A}_Q}}{dQ}&=&-\frac{d}{2}\,\Omega_d \,Q^{d-1}\ln
\left(1-\frac{b(\Lambda^2-Q^2)}{{\cal{C}}_Q(Q)}\right)     \\
\frac{d{\cal A}_Q}{dQ} &=&bQ\int_{k<Q} \frac{d^dk}{(2\pi)^d}
\frac{1}{{\cal C}_Q(k)+b(k^2-Q^2)}
\label{parasmooth}
\end{eqnarray}
where the cut-off wave vector in the smooth cut-off equation (\ref{parasmooth})
is defined as $Q\equiv \Lambda e^{-t}$. The equations hold for $Q<\Lambda$
and must be supplemented by the initial condition at $Q=\Lambda$, where ${\cal A}_Q$
reduces to the mean field expression. In both cases, as initial condition, we take a 
Landau-Ginzburg quartic form, centered around the critical density $\rho_c$, 
with coupling constants $(r,u)$ \cite{rg}:
\begin{equation}
{\cal A }_\Lambda(\rho)={\cal A }_\Lambda(\rho_c)-
r\left(\rho-\rho_c\right)^2-u\left(\rho-\rho_c\right)^4
\label{GZL}
\end{equation}
In a fully microscopic model, defined by a realistic two body interaction $w(r)$,
we expect that the quartic form (\ref{GZL}) represents the physical free energy density 
in a small neighborhood of the critical density $\rho_c$. In this case, the effective
couplings $(r,u)$ are functions of the temperature and may be estimated in mean field 
theory. Short range density fluctuations, corresponding to wave vectors $Q>\Lambda$,
are expected to renormalize the bare values, as shown in  Ref. \cite{brognara}
where this problem is examined within the HRT formalism.
The modified direct correlation function appearing in the two evolution 
equations
(\ref{parasharp}), (\ref{parasmooth}) is approximated in the spirit of the Local 
Potential Approximation of RG \cite{rg,mare}, by an Ornstein-Zernike form 
\begin{equation}
{\cal C}_Q(k) = {\cal C}_Q(0) - bk^2 = \frac{\partial^2 {\cal A}_Q}{\partial\rho^2}-bk^2
\label{oz}
\end{equation}
In Eq. (\ref{oz}) the range of the direct correlation function has been assumed to 
coincide with that of the chosen potential (\ref{phi}) with no renormalization due to
fluctuations. This is actually the only approximation introduced into the HRT formalism
and implies an analytic momentum dependence of the correlation functions in the whole phase diagram,
critical point included. In order to allow for a non vanishing critical exponent $\eta$ within HRT,
we must go beyond the parametrization (\ref{oz}) taking into account the effects of fluctuations on the range
of the correlation function. This can be accomplished by examining the second equation of the HRT
hierarchy, governing the evolution of ${\cal C}_Q(k)$ \cite{mare}.
In fact, it has been shown \cite{smooth} that, in the critical region and
in the limit $Q\rightarrow 0$, this equation allows one to reproduce
correctly the critical exponents up to the $\epsilon^{2}$ term in the
$\epsilon=4-d$ expansion, hence giving a non-vanishing value of $\eta$.
The last equality in (\ref{oz}) directly follows from the compressibility sum rule \cite{Donald}, 
according to which the structure factor evaluated at 
zero wave vector is equal to the reduced compressibility of the system.
By use of this parametrization, Eqs. (\ref{parasharp}), (\ref{parasmooth}) give rise to 
a couple of partial differential equations describing the effect of fluctuations into
the mean field free energy within the sharp and smooth cut-off formulation of HRT for a 
potential of the form (\ref{phi}). 
The resulting HRT equations are then: 
\begin{eqnarray}
\label{hrtsh}
\frac{d{\cal A}_Q}{dQ}&=& -\frac{d}{2}\,\Omega_d Q^{d-1}\ln\left [ \frac{-{\cal A}_Q^{\prime\prime}
+b\Lambda^2}{-{\cal A}_Q^{\prime\prime} +bQ^2}\right ] \\ 
\frac{d{\cal A}_Q}{dQ}&=& -\frac{b\,\Omega_d Q^{d+1}}
{-{\cal A}_Q^{\prime\prime} + b Q^2}
\label{hrtsm}
\end{eqnarray}
for the sharp and smooth cut-off respectively,
where primes mean differentiation with respect to $\rho$.
The sharp cut-off equation (\ref{hrtsh}) for a $\phi^{4}$ field
theory has already been studied in Refs. \cite{pra,max} 
both in the critical region and at phase coexistence where,
for $Q\rightarrow 0$, the term $-{\cal A}_Q^{\prime\prime} +b\Lambda^2$ in the argument
of the logarithm can be neglected. We remark that this does not affect by any
means the universal behavior at criticality or the qualitative 
features of the first-order transition discussed here.
In Ref. \cite{x} the smooth cut-off RG equations have been derived and studied for a $\phi^4$ field 
theory. The coincidence, after a trivial rescaling, of our HRT equation (\ref{hrtsm}) with 
Eq. (2.6) of Ref. \cite{x} for the special cut-off choice $n=\frac{d}{2}+1$ shows that 
indeed $i)$ the parabolic potential model defined by (\ref{phi}) is equivalent to a $\phi^4$ 
scalar field theory and $ii)$ that the smooth cut-off HRT formalism becomes 
equivalent to the RG in the scaling limit \cite{note}. 

\section {The coexistence boundary}
 
\begin{figure}[h]
\includegraphics[height=9.cm,width=8cm,angle=-90]{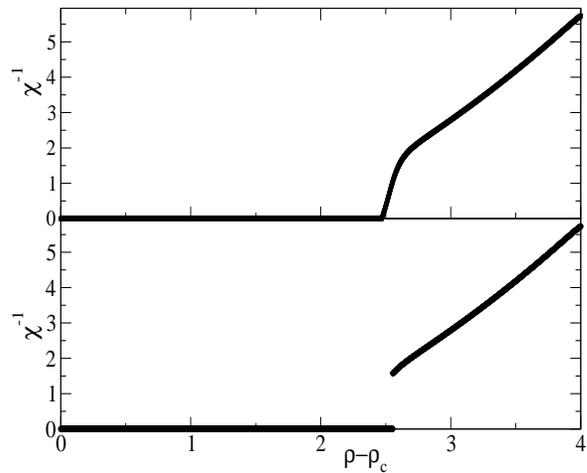}
\caption{Inverse compressibility as a function of the order parameter $\rho-\rho_c$ 
in three dimensions from the
sharp (above) and smooth cut-off formulation (below). Units are set by  
$b\Lambda^2=1$ and $\Omega_3\Lambda^3=1$, while 
the coupling constants are: $r=-0.47$ and $u=0.0035$.
Upper panel: result from the sharp cut-off formulation, 
Eq.~(\protect\ref{hrtsh}). 
Lower panel: result from the smooth cut-off formulation, 
Eq.~(\protect\ref{hrtsm}). Notice the presence of the expected discontinuity
across the coexistence boundary in the smooth cut-off case.} 
\label{both}
\end{figure}

A fully implicit predictor-corrector finite difference algorithm is used in order to solve the HRT 
evolution equations (\ref{hrtsh},\ref{hrtsm}) for a $\phi^4$ field theory. The initial condition 
(\ref{GZL}) is imposed at $Q=\Lambda$, consistently with the assumed form of the interaction 
potential (\ref{phi}). The inverse compressibility 
$\chi^{-1}=-\partial^{2}{\cal A}_{Q}/\partial \rho^{2}$ 
at the end of integration (i.e. $Q\to 0$) is 
shown in Fig.~\ref{both} as a function of $\rho-\rho_c$ 
for a representative choice of parameters in the broken-symmetry regime. 
At large densities the two formulations of HRT provide very similar results.
Moreover, both show a region
of infinite compressibility, consistent with the convexity requirement of the free energy. 
The important difference between the smooth 
and the sharp cut-off formulation is the presence of a discontinuity
across the coexistence curve in three dimensions in the smooth cut-off case, in agreement with
the expected behavior for a scalar order parameter. Instead, as already discussed in Ref. \cite{max},
the sharp cut-off HRT equation predicts the divergence of the compressibility when coexistence is
approached.

\begin{figure}
\includegraphics[height=9.cm,width=7.5cm,angle=-90]{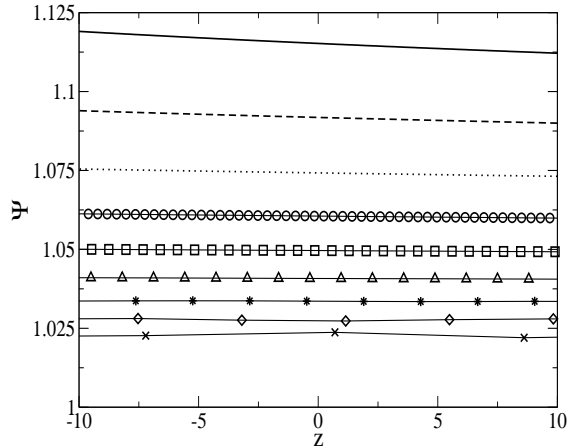}
\caption{Numerical results for the quantity $\Psi$ of Eq. \ref{logeqderiv}. Different lines correspond
to different $Q$ values. From top to bottom: $\ln Q=-4.0$ (solid line), 
$\ln Q=-4.2$ (dashed line),
$\ln Q=-4.4$ (dotted line),
$\ln Q=-4.6$ (circles), $\ln Q=-4.8$ (squares), $\ln Q=-5.0$ (triangles),
$\ln Q=-5.2$ (stars), $\ln Q=-5.4$ (diamonds),
$\ln Q=-5.6$ (crosses).}
\label{zlines}
\end{figure}

We now provide 
an analytical interpretation of the origin of the discontinuity following 
the procedure devised in \cite{max}, for the sharp cut-off formalism.
We first write our equation (\ref{hrtsm}) in a scale invariant form. By applying the rescaling:
\begin{eqnarray}
x&=&\sqrt{\frac{b}{\Omega_d}} \rho \nonumber\\
f&=&-\frac{{\cal A}_Q}{\Omega_d}
\label{rescalvar}
\end{eqnarray}
to Eq. (\ref{hrtsm}), we obtain:
\begin{eqnarray}
Q\frac{\partial f}{\partial Q}=\frac{Q^{d+2}}{f_{xx}+Q^2}
\label{eqnrescal1}
\end{eqnarray} 
where we have set $f_{xx}=\frac{\partial^2 f}{\partial x^2}$.
By performing the substitution
\begin{eqnarray}
\phi=\frac{1}{f_{xx}+Q^2} 
\end{eqnarray} 
and taking the double derivative of Eq. (\ref{eqnrescal1}) with respect to $x$, the equation
is written in quasi-linear form, suitable for an analytical study:
\begin{eqnarray}
\frac{1}{\phi^2}\frac{\partial \phi}{\partial Q} + 2Q =-Q^{d+1}\phi_{xx}
\label{eqnrescal2} 
\end{eqnarray} 
In the two-phase region, the observed convexity of the free energy implies the divergence
of the compressibility $\chi$. This means that for $Q\to 0$, $f_{xx}\propto \chi^{-1}$ 
vanishes and $\phi\to +\infty$. 
In this case, we can neglect the first term of Eq. (\ref{eqnrescal2}), so that:
\begin{eqnarray} 
\phi_{xx}=-2Q^{-d}
\label{fixx}
\end{eqnarray} 
whose solution is 
\begin{eqnarray} 
\phi=(x_0^2-x^2)Q^{-d}
\label{fi1}
\end{eqnarray} 
where the integration constant $x_0$ plays the role of rescaled coexistence density.
Indeed, this expression shows that $\phi \to +\infty$ as $Q\to 0$, consistently with our assumption,
only for $|x| < x_0$: HRT correctly predicts the existence of a finite region of infinite 
compressibility, but our analysis is not able to describe the behavior of $\phi$ across
the phase boundary, i.e. the transition between a finite solution for $|x|>x_0$ and the asymptotic form
(\ref{fi1}) inside the binodal. Following \cite{max}, it is useful to zoom-in the region close to $x_0$
by rescaling the $x$ variable as:
\begin{eqnarray}
z=(x-x_0+aQ^2)Q^{-d} 
\label{zfluct}
\end{eqnarray} 
The additive term $aQ^2$ takes into account the fluctuation corrections to the 
position of the phase boundary, which are known to play an important role
in determining the asymptotic solution on the binodal \cite{max}.
Then Eq. (\ref{eqnrescal2}) becomes:
\begin{eqnarray} 
\frac{1}{\phi^2}\left[Q\frac{\partial \phi}{\partial Q}-(dz-2aQ^{2-d}
\frac{\partial\phi}{\partial z})\right]+2Q^2\nonumber\\
=-Q^{2-d}\frac{\partial^2 \phi}{\partial z^2}
\label{lastrescal}
\end{eqnarray} 
By keeping only the dominant terms as $Q\rightarrow 0$ in Eq. (\ref{lastrescal}), 
we obtain the following fixed-point equation:
\begin{eqnarray}
\frac{2a}{\phi^2}\frac{\partial\phi}{\partial z}=-\frac{\partial^2 \phi}{\partial z^2}
\label{fpe} 
\end{eqnarray} 
Once integrated, the equation gives:
\begin{eqnarray}
\phi+\frac{2a}{c}\ln(\phi c -2a)=-cz+k
\label{eqlog}
\end{eqnarray}
where $c, k$ are integration constants. 
Inside the coexistence region, i.e. for $z\rightarrow -\infty$, 
Eq. (\ref{eqlog}) must match Eq. (\ref{fi1}) which, expressed in terms of $z$, reads 
\begin{equation}
\phi=-(x+x_0)z\approx -2x_0 z
\label{fi3}
\end{equation} 
This sets the constant $c$ at the value $c=2x_0$ . 
When the coexistence boundary is approached from outside, i.e. for $z\rightarrow +\infty$, 
Eq. (\ref{eqlog}) predicts a finite compressibility: $\phi\rightarrow 2a/c \equiv\phi_0$. 
We can then rewrite Eq. (\ref{eqlog}) as:
\begin{eqnarray}
\phi+\phi_0\ln(\phi/\phi_0-1)=-2x_0 z+ {\rm const}
\label{eqlog2}
\end{eqnarray}
which describes the asymptotic behavior of the solution across the phase boundary.
A numerical verification of such a scaling law can be obtained by considering the 
derivative of Eq. (\ref{eqlog2}) with respect to $z$:
\begin{eqnarray}
\Psi=\frac{\partial \phi}{\partial z} \frac{1}{2x_0}\frac{\phi}{\phi-\phi_0}=1
\label{logeqderiv}
\end{eqnarray}
The validity of this equation has been tested on the numerical solution of Eq. (\ref{hrtsm})
by selecting small $Q$ values ($Q < 2\times 10^{-2}$) and mesh points extremely close to the
phase boundary, so that the rescaled variable $z$ given in Eq.~(\ref{zfluct}) 
is $O(1)$. The 
results are shown in Fig.~\ref{zlines}, where several sets of data points, corresponding to
different $Q$'s and different $x$'s, are shown. The asymptotic collapse of the data towards unity,
as required by Eq. (\ref{logeqderiv}), can be appreciated. 
The results shown in the figure were obtained with a mesh 
$\Delta\rho=4\times 10^{-7}$.
Probing the asymptotic regime further would have required an even smaller 
$\Delta\rho$, so as to sample the region $z=O(1)$ for $Q$ smaller than those
considered in the figure.
The discontinuity of $\chi^{-1}$ across the binodal, shown in Fig. 1, is therefore 
a genuine consequence of the smooth cut-off HRT formalism. The analysis has been
performed for a particular choice of the potential (\ref{phi}) and for a closure of
the form (\ref{oz}), because these assumptions considerably simplify the evolution
equation (\ref{eqnrescal1}). 
We also studied the smooth cut-off HRT equation for a realistic Yukawa interaction $w(r)$,
allowing for a difference between the range of the potential and that of the
direct correlation function. The numerical solution in the asymptotic $Q\to 0$ regime shows 
the expected jump in the inverse susceptibility across the coexistence boundary. 
Our findings are consistent with the numerical analysis of Ref. \cite{x}, where the 
behavior of $\chi^{-1}$ has been investigated for a one-parameter family of smooth cut-off
RG theories which includes, as a limiting case, also the sharp cut-off form (\ref{hrtsh}), besides 
our Eq. (\ref{hrtsm}): the discontinuity is present for all values of the parameter and vanishes only
in the sharp cut-off limit. 

\section{The critical region}

\begin{figure}
\includegraphics[height=9.cm,width=8cm,angle=-90]{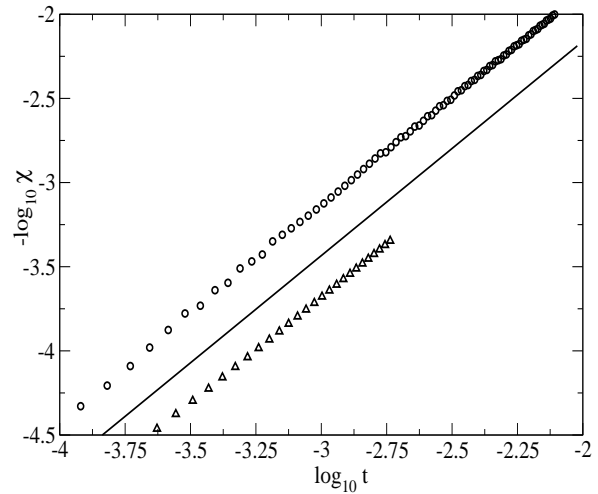}
\caption{A log-log plot of the inverse dimensionless compressibility as 
obtained by numerical integration of Eq. (\ref{hrtsm})
as a function of the reduced ``temperature" $t=|r-r_c|$. 
Triangles refer to the critical isochore, 
above the critical temperature $r>r_c$. Circles correspond to data along the 
coexistence curve for $r<r_c$. A linear fit gives an exponent $\gamma=1.28\pm 0.02$ close 
to the result obtained from the fixed point analysis $\gamma=1.300\pm 0.001$
(solid line).}
\label{figama}
\end{figure}

\begin{figure}
\includegraphics[height=9.cm,width=8cm,angle=-90]{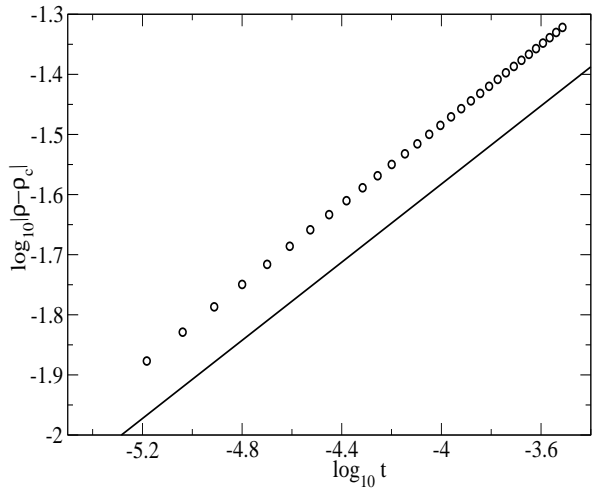}
\caption{Location of the coexistence curve as obtained  by numerical integration of Eq. (\ref{hrtsm})
as a function of the reduced ``temperature" $t=|r-r_c|$. 
A linear fit gives $\beta=0.33\pm 0.03$, consistent with 
the result from the fixed point analysis $\beta \simeq 0.325$ (solid line).}
\label{figbeta}
\end{figure}

In Section III we mentioned that the smooth cut-off HRT equation for a parabolic potential (\ref{phi}) 
reduces to a known RG structure, analogously to the sharp cut-off case \cite{mare}. 
It is therefore interesting to
evaluate the critical exponents, by studying the fixed point solution of the HRT equation and compare 
with the direct numerical solution of Eq. (\ref{hrtsm}). 
Following the usual HRT procedure \cite{pra,smooth}, inspired by the 
RG approach, we first rescale the variables 
${\cal A}_t(\rho)$ and $\rho$ (at fixed temperature $T$), in order to blow-up the critical region:
\begin{eqnarray}
H(y,Q)&=&\frac{1}{\Omega_d} \,Q^{-d}\,\left [ {\cal A}_Q(\rho_c)-{\cal A}_Q(\rho)\right ]\nonumber \\
y &=& \sqrt{\frac{b}{\Omega_d}}\, Q^{\frac{2-d}{2}}\,(\rho-\rho_c)
\label{resc}
\end{eqnarray}
where $\rho_c$ is the critical density. By performing the substitution we get:
\begin{equation}
-Q\frac{\partial H}{\partial Q}+\frac{d-2}{2}y\,H' -d\,H=\frac{1}{H_0''+1}-\frac{1}{H''+1}
\label{hrt4}
\end{equation}
where $H_0''$ is the the second derivative of $H(y)$ evaluated at the critical density. 
We first look for a fixed point solution of Eq. (\ref{hrt4})
i.e. a solution $H^*(y)$ independent of $Q$. 
According to our definition, $H(0,Q)=0$ and then also
$H^*(0)=0$. The fixed point solution 
is an even function of $y$, so that $H^{*'}(0)=0$ while $H^{*''}$ (which
explicitly appears in Eq. (\ref{hrt4})) has to be tuned in such a way that the
fixed point solution $H^*(y)$ is regular on the whole real axis.
Numerically it is more convenient to write the equation for the quantity 
$\mu(y)=H^{*'}(y)$, which can be obtained by differentiation of Eq. (\ref{hrt4})
with respect to $y$:
\begin{equation}
\frac{d-2}{2}y\,\mu' -\frac{d+2}{2}\,\mu=\frac{\mu''}{(\mu'+1)^2}
\label{fix}
\end{equation}
with initial conditions $\mu(0)=0$ and $\mu'(0)=p$. Equation (\ref{fix}) admits a single 
regular solution for a unique choice of $p$ (besides the trivial gaussian solution $\mu(y)=0$ 
which corresponds to $p=0$).  
By numerical integration of the fixed point equation (\ref{fix}) 
using a trial and error method we found a regular solution for $p=-0.186\dots$.
By setting $H(y,Q)=H^*(y)+h(y) Q^{-\lambda}$ and 
linearizing the evolution equation (\ref{hrt4}) for small $h$, we obtain the
eigenvalue equation:
\begin{equation}
\frac{d-2}{2}y\,h' +(\lambda -d)\,h=-\frac{h_0''}{(\mu_0'+1)^2}+\frac{h''}{(\mu'+1)^2}
\label{eig}
\end{equation}
where again $h_0''$ is the second derivative of the eigenfunction
evaluated at $y=0$. By shifting the eigenfunction $h(y)\to h(y)+{\rm const}$,
the constant term in Eq. (\ref{eig}) can be eliminated. The shifted function
satisfies the boundary conditions $h(0)=1$ and $h'(0)=0$. The eigenvalue $\lambda$
is determined by the requirement that $h(y)$ is free from singularities 
at all $y$'s. Relevant perturbations, driving the
system out of the critical point, correspond to positive eigenvalues. 
All the eigenfunctions can be classified according to their parity with respect to 
the symmetry $y \to -y$: Even 
solutions describe perturbations along the critical isochores, i.e. governing the temperature 
dependence of the free energy in the critical region, 
while odd eigenfunctions correspond to changes along the density axis.
As usual, a single relevant odd eigenfunction is present: $h(y)=H^{*'}(y)$ with associated eigenvalue 
$\lambda=(d-2)/2$ \cite{mare} which implies, via scaling relations, a critical exponent $\eta=0$, 
in agreement with the assumed analyticity 
of the correlation function (\ref{oz}). The numerical solution of the eigenvalue 
equation (\ref{eig}) shows that
a single relevant even eigenfunction is present, corresponding to an eigenvalue 
$\lambda=1.539\pm 0.001$ in three dimensions.
The exponent $\gamma$ governing the divergence of the compressibility 
$\chi\sim |T-T_c|^{-\gamma}$ is then given by 
$\gamma=2/\lambda = 1.300\pm 0.001$, to be compared with the accepted value for the 
Ising universality class $\gamma\sim 1.24$. 
This result agrees, within numerical uncertainties, with the estimate of $\gamma$ 
extracted from the numerical solution of the smooth cut-off HRT equation 
(\ref{hrtsm}), as shown in Fig. \ref{figama}.
Since the usual scaling relations among critical exponents are satisfied by HRT and the approximate 
Ornstein-Zernike closure (\ref{oz})
implies that the scaling exponent $\eta$ is vanishing, knowledge of $\gamma$ is 
sufficient to determine all the remaining critical exponents.
In particular, the critical exponent $\beta$  
describing the shape of the coexistence curve
($|\rho-\rho_c|\sim |T-T_{c}|^\beta$) 
is obtained by use of the scaling relations \cite{mare}, which give
$\beta=\gamma/4\simeq 0.325$. 
This value is consistent 
with the fit of the numerical results, as shown in Fig. \ref{figbeta}. 
Contrary to the prediction of HRT in the sharp cut-off formulation, 
the smooth cut-off equations lead to a
divergence of the specific heat at criticality: $C_V\sim |T-T_{c}|^{-\alpha}$. 
The critical exponent can be evaluated by
use of scaling laws with the result $\alpha=(4-3\gamma)/2\approx 0.050$. 
In Tab.~I the  values of the critical exponents in a three-dimensional fluid 
according to different approximations are reported. 
``Exact" values are derived from extrapolation of high-temperature series 
expansions \cite{field}, 
while the sharp cut-off HRT results have been 
obtained in \cite{pra}. In both HRT formulations the exponent $\eta$ vanishes 
because of the OZ assumption, and $\delta=5$ follows from scaling. 

\begin{table}
\vskip 0.2cm
\begin{tabular}{|c||c|c|c|}
\hline
Exponent &    ``Exact"    & Sharp  & Smooth \\
	 &   value   &  cut-off HRT  &  cut-off HRT  \\
\hline
$\alpha$   & 0.110 & -0.07 & 0.05   \\
\hline
$\beta$   & 0.327 & 0.345 & 0.330 \\
\hline
$\gamma$   & 1.237 & 1.378 & 1.300 \\
\hline
$\delta$    & 4.789 & 5 & 5\\
\hline
$\eta$ & 0.036 & 0 & 0 \\
\hline
\end{tabular}
\caption{HRT estimates of the critical exponents for the Ising universality class in three dimensions 
compared to the exact values \cite{field} obtained by extrapolation 
of high-temperature series expansions. Sharp cut-off HRT results 
from Ref.~\protect\cite{mare}}
\end{table}

\section{Conclusions}

The smooth cut-off HRT equation (\ref{hrt}) has been derived in the framework of liquid state theory 
for a general fluid model. The theory has been then specialized to a ``parabolic potential"
which corresponds to a microscopic realization of a $\phi^4$ theory. The resulting partial
differential equation turns out to be equivalent to a particular formulation of smooth cut-off RG 
theory \cite{x}. We stress that, contrary to other RG methods, the gradual turning 
on of fluctuations which characterizes the HRT approach, is performed directly 
on the physical quantities, like the free energy and
no {\sl a priori} mapping to effective models is required. This allows to preserve the 
information on both the universal and the non-universal properties of the system. 
The equation has been numerically solved both above and below the critical temperature 
showing that smooth cut-off HRT provides an extremely promising tool for the description of 
the phase diagram of fluids. Among the unique features of this formalism we recall
\begin{itemize}
\item A treatment of the critical region consistent with scaling laws and characterized by
non classical critical exponents in good agreement with the exact values.
\item A built-in mechanism leading to the convexity of the free energy which implements Maxwell's
construction via the inclusion of long range density fluctuations.
\item An accurate description of the first order transition, which predicts the correct jump
of the inverse compressibility at the phase boundary. 
\end{itemize}
Although in this paper we used the smooth cut-off HRT for the analysis of a coarse-grained hamiltonian, the
$\phi^4$ field theory, Eq. (\ref{hrt}) can be directly applied to fully microscopic 
models of fluids, as already shown in the sharp cut-off case (see \cite{mare} and references therein). 
A particularly favorable choice may be the Yukawa fluid, which allows for an
exact implementation of the core condition for the whole sequence of intermediate models (\ref{smoothtail}) 
interpolating between the reference and the fully interacting systems.
This feature can make the smooth cut-off formulation a valuable tool for the 
theoretical investigation of the phase behavior of simple and complex fluids.

\section{Acknowledgment}
We acknowledge support from the Marie Curie program of the European Commission,
contract number MRTN-CT2003-504712.


\begin{thebibliography}{10}
\bibitem{note} The equivalence between the asymptotic {\sl sharp cut-off} 
HRT equations and an implementation of RG for the $\phi^4$ theory has been
already proved in Refs. \cite{pra}.
\end{thebibliography}
\end{document}